\documentclass[iop,revtex4]{emulateapj}
\usepackage{times}

\usepackage{natbib}
\usepackage{amsmath,bm}
\usepackage{multirow}
\usepackage{enumerate}
\usepackage{comment}
\usepackage{url}
\usepackage[colorlinks, citecolor=blue]{hyperref}

\begin{document}

\shorttitle{Double-Decker Filament II} %

\shortauthors{Kliem et al.}

\title{Slow Rise and Partial Eruption of a Double-Decker Filament.\\ 
       II Modeling by a Double Flux Rope Equilibrium}

\author{Bernhard Kliem\altaffilmark{1,2,3,4},
        Tibor T\"{o}r\"{o}k\altaffilmark{5},
        Viacheslav S. Titov\altaffilmark{5},
        Roberto Lionello\altaffilmark{5}, \\ 
        Jon A. Linker\altaffilmark{5},
        Rui Liu\altaffilmark{6,7},
        Chang Liu\altaffilmark{7}, and
        Haimin Wang\altaffilmark{7}}

\altaffiltext{1}{Institute of Physics and Astronomy, University of
                 Potsdam, Karl-Liebknecht-Str.~24, 14476 Potsdam, Germany}
\altaffiltext{2}{Yunnan Observatories, Chinese Academy of Sciences,
                 Kunming 650011, China}
\altaffiltext{3}{Mullard Space Science Laboratory, University College London,
                 Holmbury St.~Mary, Dorking, Surrey RH5 6NT, UK}
\altaffiltext{4}{College of Science, George Mason University, Fairfax, VA
                 22030, USA}
\altaffiltext{5}{Predictive Science, Inc., San Diego, CA 92121, USA}
\altaffiltext{6}{CAS Key Laboratory of Geospace Environment, University of
                 Science and Technology of China, Hefei 230026, China}
\altaffiltext{7}{Space Weather Research Laboratory, Center for
                 Solar-Terrestrial Research, NJIT, Newark, NJ 07102, USA}

\email{bkliem@uni-potsdam.de}

\journalinfo{Manuscript doubledeck2\_v.b2.tex}  
\submitted{Submitted: May 29, 2014}  

\begin{abstract}  Force-free equilibria containing two vertically
arranged magnetic flux ropes of like chirality and current direction
are considered as a model for split filaments/prominences and
filament-sigmoid systems. Such equilibria are constructed analytically
through an extension of the methods developed in \citet{td99} and
numerically through an evolutionary sequence including shear flows,
flux emergence, and flux cancellation in the photospheric boundary. It
is demonstrated that the analytical equilibria are stable if an
external toroidal (shear) field component exceeding a threshold value
is included. If this component decreases sufficiently, then  both flux
ropes turn unstable for conditions typical of solar active regions,
with the lower rope typically being unstable first. Either both flux
ropes erupt upward, or only the upper rope erupts while the lower rope
reconnects with the ambient flux low in the corona and is destroyed.
However, for shear field strengths staying somewhat above the
threshold value, the configuration also admits evolutions which lead
to partial eruptions with only the upper flux rope becoming unstable
and the lower one remaining in place. This can be triggered by a
transfer of flux and current from the lower to the upper rope, as
suggested by the observations of a split filament in Paper~I
\citep{Liu&al2012}. It can also result from tether-cutting
reconnection with the ambient flux at the X-type structure between the
flux ropes, which similarly influences their stability properties in
opposite ways. This is demonstrated for the numerically constructed
equilibrium.

\end{abstract}


\section{Introduction}\label{s:intro}

\begin{figure*}
\centering \includegraphics[width=.75\linewidth]{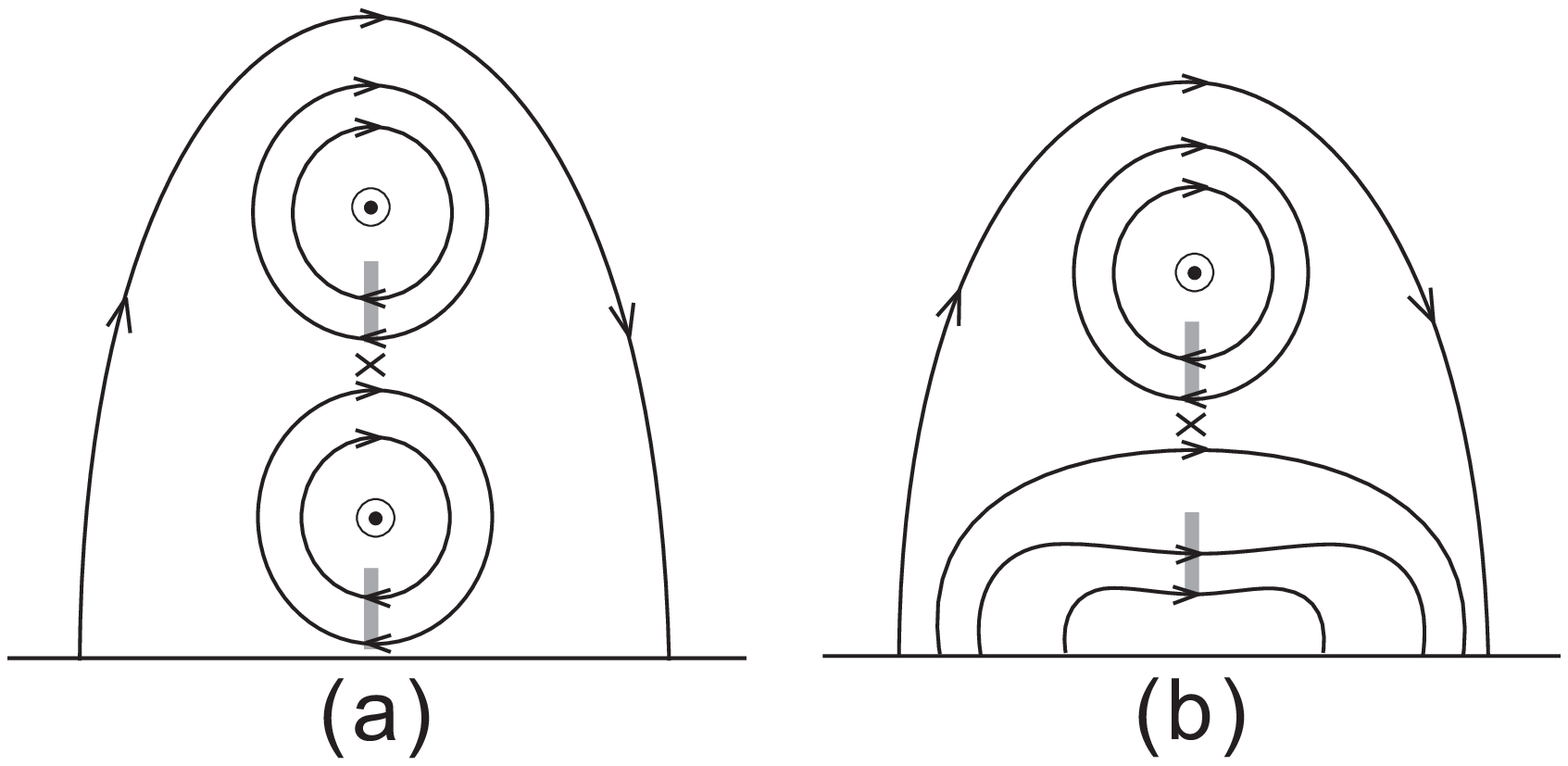}  
\caption{\label{f:cartoon_w_HFT}
 Cartoon illustrating the cross section of the two double-decker
 filament configurations suggested in Paper~I:
 (a) double flux rope equilibrium;
 (b) flux rope above a sheared arcade.
 The axial field of both filament branches points in the same
 direction (out of the plane in the specific case analyzed in Paper~I).
 The HFT is indicated by a small cross.
 Slabs of gray color indicate the filament body trapped in dipped
 field line sections.}
\end{figure*}

The magnetic structure of solar prominences (filaments if observed on
the disk) is one of the major debated subjects in solar physics: a
flux rope and a sheared loop arcade are being controversially
discussed \cite[e.g.,][]{Mackay&al2010}. An extension of the flux rope
concept is suggested to be relevant in some cases by the analysis of a
``double-decker'' filament in \citet{Liu&al2012}, hereafter Paper~I.
The filament, located in active region (AR) 11093, consisted of two
main branches and experienced a partial eruption, ejecting only the
upper branch into a coronal mass ejection (CME), on 2010 August~7. The
clear vertical separation of the filament branches prior to eruption
and the stability of the lower branch in the course of the eruption
both suggest that the filament may have formed in a double flux rope
structure. An alternative explanation in terms of a single flux rope
situated above an arcade which contains the lower filament branch is
also possible, but has the disadvantage that basically different
magnetic structures for the two filament branches are implied. These
two configurations are illustrated in Figure~\ref{f:cartoon_w_HFT},
and their possible formation mechanisms and relevance for the event
considered in Paper~I will be discussed in Section~\ref{s:discussion}.
The formation of both filament branches within a single flux rope or
within an arcade is far less likely, as it requires two special
conditions to be satisfied simultaneously: the trapping of filament
material at two clearly separated heights and the formation of the
flare current sheet in the course of the eruption at an intermediate
height.

A striking phenomenon observed during the slow-rise phase of the upper
filament branch in AR~11093 prior to its eruption was the transfer of
material from the lower to the upper branch, occurring in several
episodes. Assuming dominantly horizontal field orientation in the
filament, a corresponding transfer of flux is implied. This may have
caused the eruption by producing an imbalance between the flux in the
upper branch and the ambient flux \cite[e.g.,][]{Su&al2011}, lifting
the upper branch into the torus-unstable height range \citep{kt06,
Kliem&al2013}.

\citet{Zhu&Alexander2014} describe a very similar event in AR~11475 on
2012 May~9--10. The filament consisted of two branches separated in
height. Several discrete episodes of mass transfer from the lower to
the upper branch occurred in the two days preceding the eruption of
the upper branch into a CME, while the lower branch remained stable.

A double flux rope configuration was also suggested to exist prior to
an eruption in AR~11520 on 2012 July~12 \citep{XCheng&al2014}. The
event featured an erupting sigmoidal ``hot channel'' structure above a
stable filament, similar in many respects to an eruption on 2010
August~1 investigated in \citet{liu10}. Since this combination is not
rare \cite[e.g.,][]{pevtsov02, liu07, liu08}, the double flux rope
structure may be more common than expected so far.

Employing two magnetohydrodynamic (MHD) modeling approaches, the
present paper substantiates the suggestion that a double flux rope
configuration can be consistent with the long-term stability of the
filaments studied in Paper~I, \citet{Zhu&Alexander2014}, and
\citet{XCheng&al2014} during their slow rise phase, and with the
partial eruptions. This also introduces new scenarios for partial
filament eruptions. For some events these may be an alternative to the
dynamical splitting and partial expulsion of a single, originally
purely O-type flux rope whose top part has become unstable while the
bottom part remains line-tied in the photosphere \citep{gf06}. In such
a configuration the flux splits only after the main acceleration of
the ejection has commenced.

We restrict the consideration to the case that the flux ropes are of
like chirality, with the axial field component pointing into the same
direction, as suggested by the relevant events quoted above. This can
be expected to be the typical situation in split filaments that form
side by side in the same filament channel and have common or close end
points. The axial (toroidal) currents in the ropes then point in the
same direction implying an attractive force between them.

The existence of a stable double flux rope equilibrium is not trivial
if the ropes are relatively close to each other so that the force
between them is comparable to or larger than the force exerted by the
ambient field. Perturbations will then trigger a pinching of the
X-type magnetic structure between the ropes, which in general is a
hyperbolic flux tube \cite[HFT; see][]{titov02}. The subsequent
reconnection at the formed current sheet will start redistributing the
magnetic flux between the ropes and the ambient field. This may cause
the perturbation to grow, eventually merging the ropes or pushing them
apart, depending on the properties of the configuration and
perturbation.

Using an extension of the Shafranov equilibrium of a single toroidal
force-free flux rope \citep{Shafranov1966, td99}, we first demonstrate
the existence, stability, and instability of an equilibrium containing
two vertically arranged force-free flux ropes in bipolar external
field (Section~\ref{s:equilibrium}). We find that the external field's
toroidal (shear) component is a key parameter controlling the
stability of the configuration if the ropes are relatively close to
each other. We also demonstrate that a scenario of current and flux
transfer from the lower to the upper rope, which resembles the flux
transfer indicated by the observations in Paper~I and
\citet{Zhu&Alexander2014}, leads to instability of the upper flux rope
only.

In Section~\ref{s:splitting_FR} we describe an MHD simulation which
exhibits the formation of a split flux rope during the slow-rise phase
of a modeled filament eruption. Different from the model by
\citet{gf06}, this timing corresponds to the observations of the
double-decker filaments presented in Paper~I and in
\citet{Zhu&Alexander2014}.

\section{Stability and Instability of a Double Flux Rope Equilibrium}
\label{s:equilibrium}

\subsection{Construction of the Equilibrium}
\label{ss:construction}

We build on the construction of an approximate analytical equilibrium
of a toroidal force-free flux rope in bipolar current-free external
field described in \citeauthor{td99} (\citeyear{td99}, hereafter
TD99), done in two steps. First, the ``external equilibrium'' of the
rope in a given simple (axisymmetric) external poloidal field
$B_\mathrm{ep}$ is determined by balancing the Lorentz force of the
flux rope current in the field $B_\mathrm{ep}$ with the Lorentz
self-force (hoop force) of the current. Then an approximate ``internal
equilibrium'' of the current channel in the core of the rope is
constructed by matching the expressions for a straight force-free
current channel in each cross section of the toroidal channel to the
external field at its surface.

A second, larger flux rope, lying in a concentric arrangement in the
same plane as the first rope, can easily be added at a sufficiently
large distance, such that its influence on the external equilibrium of
the first rope is negligible. The second rope can then be constructed
in the same way as a single rope, except that the known poloidal field
of the first rope must be added to the external poloidal field in
determining the external equilibrium of the second rope. This very
simple approximation yields equilibria that readily relax to a
numerical equilibrium very near to the analytical one if the ratio of
the major radii is not smaller than $\approx\!4$.

In order to find equilibria of two flux ropes with smaller distance,
the construction of the external equilibrium in TD99 can
straightforwardly be generalized (see below). Doing so also for the
internal equilibrium is a very involved task, which we do not aim to
pursue here. However, applying the expressions for the internal
equilibrium without modification to each channel individually yields
an acceptable approximation down to ratios of the major radii of
$\approx\!2.5$. Subsequent numerical MHD relaxation quickly adjusts
the internal equilibrum of the ropes and, supported by line tying in
the photospheric boundary, yields a nearby force-free numerical
equilibrium in the stable part of the parameter space.

We now consider the external equilibrium of two concentric toroidal
current channels lying in a plane. Here the relevant poloidal field is
the superposition of the external poloidal field, $B_\mathrm{ep}$, and
the poloidal field by the other current channel, $B_I$, where $I$ is
the total toroidal (ring) current of the channel. In TD99 the external
poloidal field is due to a pair of auxiliary magnetic charges ${\pm}q$
on the symmetry axis of the torus at distance ${\pm}L$ from the torus
plane (or simply the field connecting the corresponding ``sunspots''
in the photospheric plane); we will also refer to this field component
as $B_q$. Using subscripts 1 and 2 to denote quantities of the inner
(lower) flux rope FR1 and outer (upper) flux rope FR2, respectively,
we have the following dependencies on the flux rope currents:
$F_{I_{1,2}}\propto I_{1,2}^2$ for the Lorentz self-force;
$F_{q_{1,2}}\propto I_{1,2}$ for the force by the field from the
magnetic charges; and $F_{B_{1\leftrightarrow2}}\propto I_1I_2$ for
the force due to the field of the other current channel. The resulting
force equations,
\begin{eqnarray}
0&=&F_{I_1}+F_{q_1}+F_{B_{2\to1}}=b_1I_1^2+c_1I_1+e_1I_1I_2    \label{e:1}\\
0&=&F_{I_2}+F_{q_2}+F_{B_{1\to2}}=b_2I_2^2+c_2I_2+e_2I_1I_2\,, \label{e:2}
\end{eqnarray}
can easily be solved for $I_1$ and $I_2$, given the geometry
($R_{1,2}$, $a_{1,2}$, $d$, $L$) and the strength of the field $B_q$,
set by $q$ and $L$. Here $R$ and $a$ denote major and minor torus
radius, respectively, and $d$ is the depth of the torus center below
the photospheric plane. The expressions of the coefficients $b_{1,2}$,
$c_{1,2}$, and $e_{1,2}$ can be found in TD99 (their Equations 5, 4,
and 25, respectively). As in TD99, an axisymmetric external toroidal
field $B_\mathrm{et}$ of arbitrary strength can be superposed. It is
provided by an auxiliary line current $I_0$ running along the symmetry
axis of the tori. Given the geometry, $q$, and the currents $I_0$,
$I_1$, and $I_2$, the equilibrium field can straightforwardly be
computed as the superposition of $B_q$, $B_\mathrm{et}$, $B_{I_1}$,
and $B_{I_2}$, using the expressions (16), (20), and (31) in TD99.

The resulting equilibrium is only a very crude approximation of the
suggested interpretation of the double-decker filament in Paper~I in
terms of two flux ropes. Both ropes in the model must have toroidal
shape, so that their footpoints are quite separate, different from the
observed configuration. Moreover, the use of a line current as the
source of the external toroidal field, dictated by the concentric
arrangement of the tori, prevents us from realistically modeling an
ejective eruption (a CME). Since the resulting $B_\mathrm{et}$ falls
off only linearly with distance from the torus center, it enforces any
eruption to remain confined for realistic values of its strength at
the position of the flux ropes \citep{Roussev&al2003, tk05}. On the
other hand, the configuration allows us to demonstrate the existence
of the suggested equilibrium, the instability of only the upper flux
rope for certain parameter settings, and the key role of
$B_\mathrm{et}$. This can be done for a geometry that matches the
observed height relationships between the lower and upper filament
branches at the apex points of the two flux ropes.

The double flux rope equilibrium is intrinsically less stable than the
equilibrium of a single toroidal flux rope, due to the additional
force between the ropes. To attain a stable force-free equilibrium in
the absence of an external toroidal field, the ropes must be
positioned sufficiently far apart, so that $F_{B_{1\leftrightarrow2}}$
is small compared to $F_{q_{1,2}}$, and $B_q$ must be relatively
uniform, i.e., $L$ must be large. The slope of the total poloidal
field as a function of $R$ then remains sufficiently small at the
positions of both ropes, so that both are stable against vertical
displacements \citep{kt06}. Here we have to consider a situation with
the flux ropes situated relatively close to each other and the sources
of the field $B_q$ also being relatively close (see below). In this
case, an external toroidal field is required for stability. If one or
both ropes are displaced by a perturbation, the compression of this
field component will counteract the perturbation. Thus, the strength
of the external toroidal field is a key parameter deciding between
stability and instability of the configuration.

This is confirmed in the following subsection by MHD relaxation runs,
which also show that a size ratio $R_2/R_1\gtrsim2.5$ is required for
the analytical equilibrium to be close to a stable numerical one with
the HFT lying not too close to the current channel in the lower rope.
In the stable domain of parameter space, reconnection at the HFT
remains very weak, just redistributing the fluxes to the extent needed
to reach the nearby numerical equilibrium.

Prominence material in flux ropes is supposed to be trapped in dips of
the field lines, i.e., it can occupy a slab-like volume extending
between the bottom flux surface and the magnetic axis of the rope
(Figure~\ref{f:cartoon_w_HFT}). The height measurements of the two
filament branches on 2010 August~7 (Section~2.3 in Paper~I) thus
suggest apex heights of 12 and 36~Mm for the magnetic axis of FR1 and
FR2, respectively, and an apex height of 25~Mm for the HFT (which is
the bottom of the upper flux rope on the vertical line through the
apex). The first two measurements are met, for example, by the choice
$R_1=16$~Mm, $R_2=40$~Mm, $d=4$~Mm. The third measurement can be met
to a good approximation by the choice $a_1=4$~Mm, $a_2=6$~Mm,
resulting in an HFT apex height of 23~Mm. We have chosen the current
channels to be relatively thin, so that both possess a large aspect
ratio, which guarantees relatively high precision in the construction
of the equilibrium (TD99). Note that it is the larger cross section of
the magnetic structure (the flux rope) which matters for the location
of field line dips, not the cross section of the current channel.  To
have the HFT apex lying exactly at the estimated height of 25~Mm, we
would have to choose $a_2=2$~Mm, smaller than $a_1$, which we consider
neither appropriate nor necessary for the purpose of this study. Under
the force-free constraint, the relatively small values of the minor
radii imply relatively high values of the flux rope twists. Both ropes
are stabilized against the helical kink mode by the external toroidal
field. We set $L=8$~Mm, corresponding to the observed distance of the
main photospheric flux concentrations near the middle section of the
filament studied in Paper~I (see Figure~2 in Paper~I).

\subsection{Numerical Simulations}\label{ss:simulations}

\subsubsection{Stable Configuration}\label{sss:stable}

The resulting analytical equilibrium is used as initial condition in
zero-beta MHD simulations in a cubic Cartesian box more than five
times higher than FR2 and resolving the minor diameter of the current
channel in FR1 by 35 grid cells. The initial density is specified as
$\rho_0(\bm{x})=|\bm{B}_0(\bm{x})|^{3/2}$, where $\bm{B}_0(\bm{x})$ is
the initial magnetic field, so that the Alfv\'en velocity decreases
slowly with height above the flux rope, as in the solar corona.

\begin{figure}
\centering \includegraphics[width=.83\linewidth]{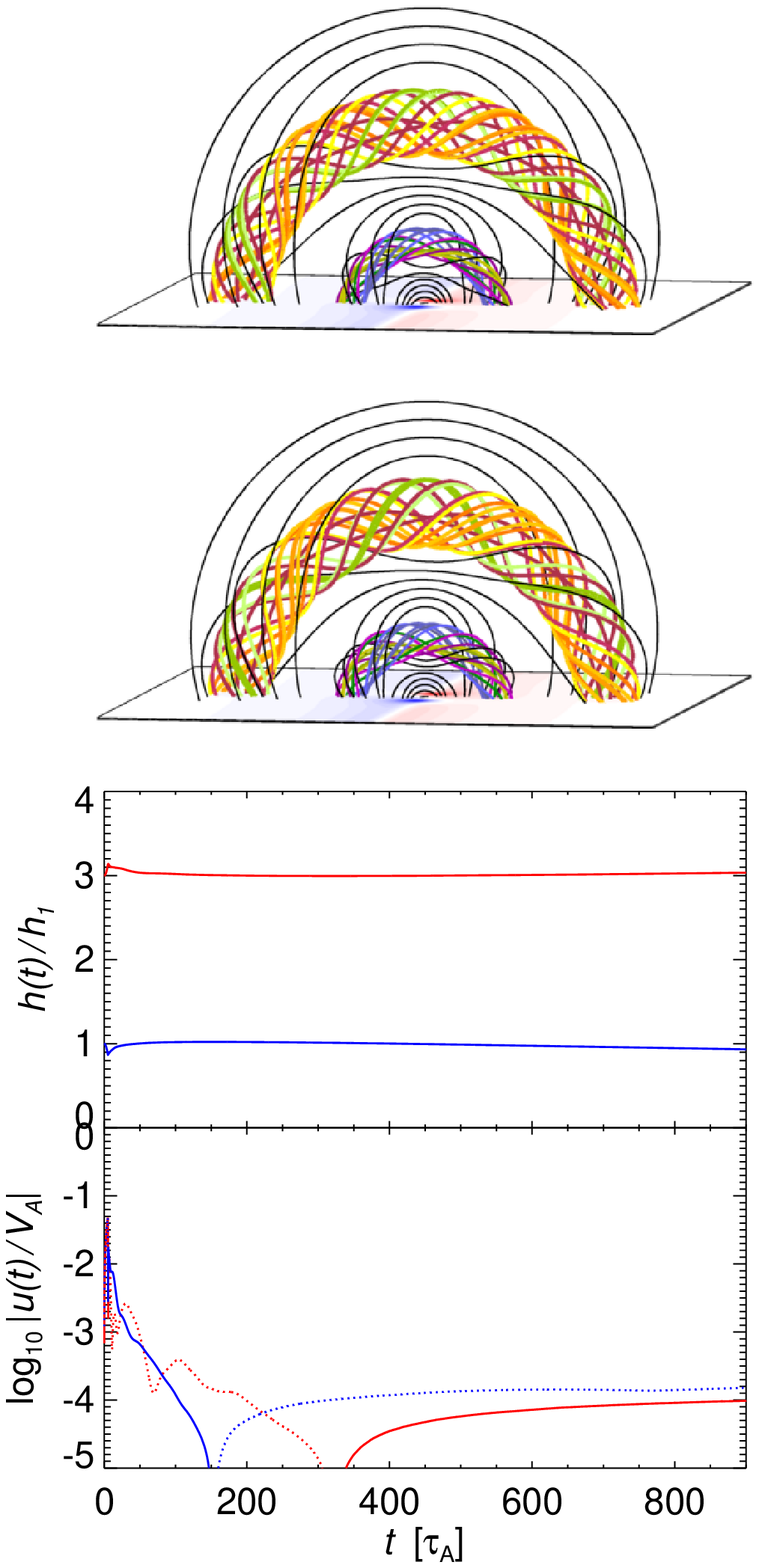}   
\caption{\label{f:relax}
 Field lines showing the two flux ropes in the stable analytical
 equilibrium with $B_\mathrm{et}=1.7B_q$, slightly above the marginal
 stability value of the external toroidal field (top panel) and after
 numerical relaxation at $t=645\tau_\mathrm{A}$ (middle panel). The
 field lines lie in flux surfaces near the surface of the current
 channels (slightly inside the channel for FR1).
 All black field lines pass through the vertical axis in the middle of
 the system, illustrating the apex heights of the two HFTs (below the
 lower rope FR1 and between FR1 and FR2) by the transition between
 downward and upward concave curvature.
 The magnetogram, $B_z(x,y,0,t)$, is displayed in the bottom plane.
 The bottom panel shows height and velocity of the fluid elements at
 the apex points of the magnetic axes of the two ropes in the
 relaxation run. Due to the symmetry of the system, these fluid
 elements move only vertically. The relaxation run includes an initial
 velocity perturbation in small spherical volumes centered at the two
 apex points, applied up to $t=5~\tau_\mathrm{A}$. Downward velocities
 are shown dotted.}
\end{figure}



To check the equilibrium currents $I_{1,2}$ obtained from
Equations~(\ref{e:1}--\ref{e:2}), the field $B_\mathrm{et}$ is first
set to a value somewhat (about 20~percent) below the value that
provides stability. After normalizing by the apex height of FR1,
$h_1=R_1-d$, and by the corresponding Alf\'en time
$\tau_\mathrm{A}=h_1/V_\mathrm{A}$, where $V_\mathrm{A}$ is the
Alfv\'en velocity at the magnetic axis of FR1, the system is
integrated in time for $100~\tau_\mathrm{A}$. This reveals both the
quality of the equilibrium and its unstable nature. Within the first
$\sim30~\tau_\mathrm{A}$, the system attempts to settle to an exact
numerical equilibrium from the approximate analytical one, with the
velocities reaching a small peak and then falling to the very modest
values of only $\approx\!0.002~V_\mathrm{A}$ in FR1 and
$\approx\!-0.001~V_\mathrm{A}$ in FR2, which indicates that the
analytical equilibrium is nearly perfect.  Subsequently, the
velocities of both ropes begin to increase very gradually. This
increase continues (roughly doubling the minimum values of the
velocity by the end of the run), which indicates instability.
By slightly modifying one of the currents, a nearly perfect
equilibrium is found, with the residual velocities of FR1 falling
monotonically to $1.4\times10^{-4}~V_\mathrm{A}$ and the ones of FR2
oscillating very gradually around a value
$6\times10^{-4}~V_\mathrm{A}$ by $t=100~\tau_\mathrm{A}$; this
requires reducing $I_1$ by 0.5\%.


Next, this configuration is integrated for a range of $B_\mathrm{et}$
values, with a velocity perturbation applied at the apex of both flux
ropes to find the minimum stabilizing external toroidal field. The
velocity perturbation is applied for $5~\tau_\mathrm{A}$, linearly
ramped up to the peak value of $\pm\,0.05~V_\mathrm{A}$ and then
switched off. Marginal stability (the critical value
$B_\mathrm{et,\,cr}$) is found for $B_\mathrm{et}$ slightly below
$1.7B_q$, with $B_\mathrm{et}$ and $B_q$ taken at the lower flux
rope's magnetic axis. Figure~\ref{f:relax} shows field lines of the
stable analytical equilibrium with $B_\mathrm{et}=1.7B_q$ and of the
configuration after numerical relaxation, along with height and
velocity of the fluid elements at the apex points of the two flux
ropes in the relaxation run. This stable numerical equilibrium is very
close to the analytical one.

The external toroidal (shear) field strength required for stability is
relatively high, exceeding the poloidal component of the external
field (perpendicular to the filament). This situation can be realized
on the Sun if the filament ends in the main flux concentrations of the
active region. These sources then provide not only the axial field of
the filament, but also give the ambient field a strong component in
the direction of the filament. The filament investigated in Paper~I
did have this configuration.


\subsubsection{Unstable Configurations}\label{sss:unstable}

\paragraph{Full Eruptions}\label{p:full_eruptions}


If the stabilizing external toroidal field strength is slightly
reduced below the threshold value $B_\mathrm{et,\,cr}\approx1.7B_q$,
then the configuration can no longer be relaxed to a nearby
equilibrium (Section~\ref{sss:stable}). The nature of the instability
and the complexity inherent in the configuration become apparent when
$B_\mathrm{et}$ is reduced considerably. In this and the following two
paragraphs we refer to simulations with
$B_\mathrm{et}=B_\mathrm{et,\,cr}/3$ when quoting numbers. Identical
qualitative behavior is obtained in the range
$0\le B_\mathrm{et}\lesssim B_\mathrm{et,\,cr}/2$. The simulations
confirm that both flux ropes are unstable against vertical
displacements (i.e., the torus instability) for the geometrical
parameters chosen. The relevant parameter is the ``decay index'' of
the total poloidal field at the position of each rope,
$n=-d\ln{(B_q+B_I)}/d\ln{h}$, where $B_I$ is the poloidal field from
the other rope. Its threshold value lies in the range
$n_\mathrm{cr}\approx1.5\mbox{--}2$ if $B_\mathrm{et}$ is small
\citep{kt06, tk07}. Our configuration yields $n_1=3.1$ and $n_2=2.8$,
implying a higher growth rate for the instability of the lower flux
rope FR1. (Additionally, higher Lorentz forces can be expected to
develop in the evolution of the lower rope, since the field strength
and current density are higher.) The decay index at the position of
the lower flux rope generally has a high value, since the poloidal
field of the upper flux rope, $B_{I_2}$, and the external poloidal
field, $B_q$, are oppositely directed under the upper rope. The decay
index at the position of the upper flux rope is largely determined by
the field $B_q$, which has a supercritical decay index at heights
exceeding $\sim\!L$. This condition is clearly met by our choice of
geometrical parameters. Obviously, both flux ropes in a double flux
rope equilibrium of the type studied in this paper tend to be torus
unstable if the external toroidal (shear) field falls below the
threshold value $B_\mathrm{et,\,cr}$.

\begin{figure}
\centering \includegraphics[width=.77\linewidth]{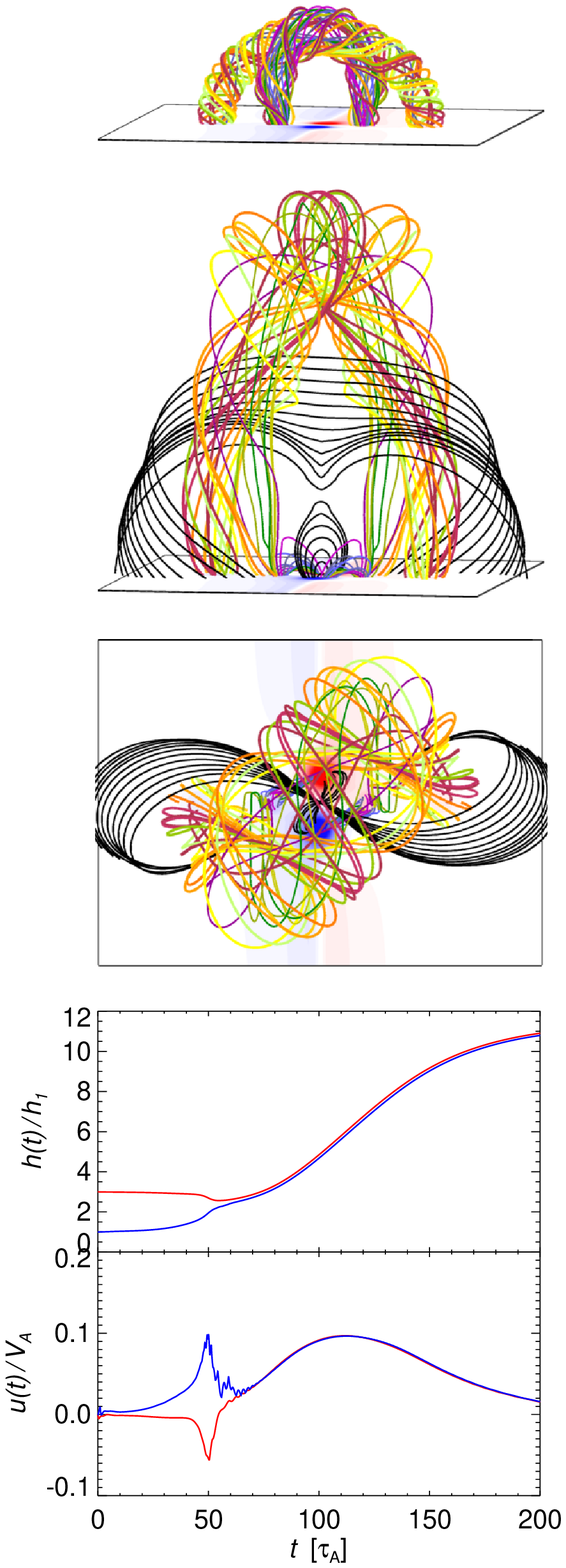} 
\caption{\label{f:FR1up}
 Instability and full eruption of the double flux rope equilibrium in
 the case of sub-critical external toroidal field,
 $B_\mathrm{et}=B_\mathrm{et,\,cr}/3$, and upward motion of the lower
 rope. Field line plots similar to Figure~\ref{f:relax} are shown at
 $t=65~\tau_\mathrm{A}$ (top panel) and $t=165~\tau_\mathrm{A}$
 (second and third panel). The lower flux rope largely merges with the
 upper one; the other part of its flux reconnects with the ambient
 flux to join the forming post-eruption arcade.
 The motion of the fluid elements at the apex points of the flux ropes
 is displayed in the bottom panel. A small upward initial velocity
 perturbation is applied at the apex of the lower flux rope.}
\end{figure}

\begin{figure}
\centering \includegraphics[width=.77\linewidth]{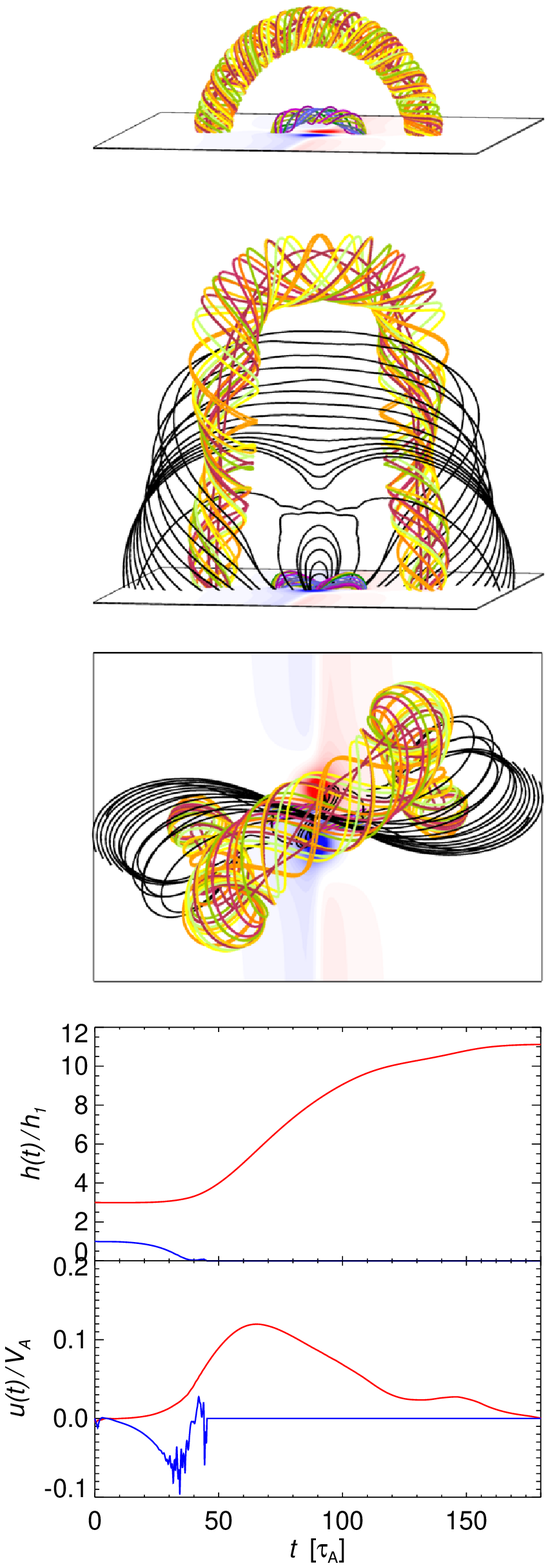}  
\caption{\label{f:FR1down}
 Same as Figure~\ref{f:FR1up} for the case of downward moving lower
 rope (following a small downward velocity perturbation) at
 $t=30~\tau_\mathrm{A}$ (top panel) and $t=100~\tau_\mathrm{A}$
 (second and third panel). Here all of the flux in the lower rope
 moves downward and reconnects with the ambient flux.}
\end{figure}


Two different evolutions are enabled by the dominant instability of
the lower flux rope. When a small upward perturbation is applied to
FR1, it then shows an exponential rise which saturates as the upper
rope FR2 is approached. FR2 stays near its initial position in this
period. Subsequently, the ropes merge, forming an arch which expands
upward with a velocity of order $0.1~V_\mathrm{A}$
(Figure~\ref{f:FR1up}). The rise is terminated at about four times the
initial height of the upper rope, $h_2=R_2-d$, by the onset of
reconnection between the legs of the strongly writhing rope and the
ambient field. The writhing is here primarily due to the presence of
$B_\mathrm{et}$ \citep{if07, Kliem&al2012}, i.e., it does not indicate
the development of the helical kink instability. The velocity doubles
and the rise continues until the upper boundary of the box is
approached if $B_\mathrm{et}$ is reduced further, below
$\sim B_\mathrm{et,\,cr}/10$. This full eruption is clearly driven by
the stronger torus instability of the lower flux rope and is very
similar to the eruption of a single torus-unstable flux rope.


When a small downward perturbation is applied to FR1, it shows a
short, exponentially increasing downward displacement until the bottom
boundary of the box (the model photosphere) is hit. FR1 then
reconnects with the sunspot field, splitting in two low-lying ropes
which come to rest at the bottom of the box. FR2 immediately begins an
exponential rise which enters the saturation phase at about $1.5h_2$,
followed by an approximately linear rise, again with a velocity of
order $0.1~V_\mathrm{A}$ (Figure~\ref{f:FR1down}). Similar to the case
in Figure~\ref{f:FR1up}, reconnection between the legs of the writhing
rope and the ambient field terminates the rise at $\sim\!4h_2$, but a
further reduction of $B_\mathrm{et}$ to $\sim B_\mathrm{et,\,cr}/10$
allows the upper flux rope to double the rise velocity and to escape
(i.e., reach the top boundary of the box). Hence, a partial eruption
(of only the upper flux rope) occurs, but it is accompanied by a
strong change of the lower flux rope.

The scenario of decreasing $B_\mathrm{et}$ considered here may easily
be realized on the Sun as the sources of the external toroidal field
weaken by flux dispersal and cancellation. Thus, the eruption of
double flux rope equilibria on the Sun will often involve a complete
change of the configuration, both for full and partial eruptions.
Since $B_\mathrm{et}$ will typically decrease only very gradually, the
instability will set in long before a value of order
$B_\mathrm{et,\,cr}/3$ is reached. Nevertheless, the upward directed
velocities in such eruptions can be expected to be similar to or even
higher than the values found in our simulations, since the external
toroidal field in the corona is expected to fall off with height above
the filament much faster than the model field, which is unrealistic in
this regard (Section~\ref{ss:construction}). The downward directed
velocities should remain considerably below the simulated ones, since
$B_\mathrm{et}$ will be much closer to the threshold value in this
height range. Thus, in the scenario of eruptions enabled by gradually
decreasing external toroidal field, a partial eruption may have a less
dramatic effect on the lower flux rope than found in the simulation,
but a destruction of the lower flux rope must still be expected.
However, partial eruptions involving only the upper flux rope, with
the lower flux rope remaining stable at its initial position, are also
possible.

\paragraph{Partial Eruptions with a Stable Lower Flux Rope}
\label{p:partial_eruptions}


A partial eruption can result under the same scenario of decreasing
$B_\mathrm{et}$ if the upper flux rope is sufficiently twisted, so
that the helical kink instability develops in the upper rope while the
lower rope is still stable against the torus instability. We have
verified this possibility by reducing the minor radius of the upper
rope to $a_2=3.5$~Mm, which doubles its twist to about $10\pi$, with
$B_\mathrm{et}$ kept at the critical value of $1.7B_q$ of the original
configuration. The upper rope then kinks upward, while the lower rope
stays very close to its original position with very small residual
velocity for more than $100~\tau_\mathrm{A}$, i.e., apparently in a
stable state. However, the occurrence of such high twist values is
very unlikely and has so far been reported only in a single case
\citep{rcz03}. Therefore, we now consider another scenario for partial
eruptions of double flux rope equilibria.

\begin{figure*}
\centering \includegraphics[width=.66\linewidth]{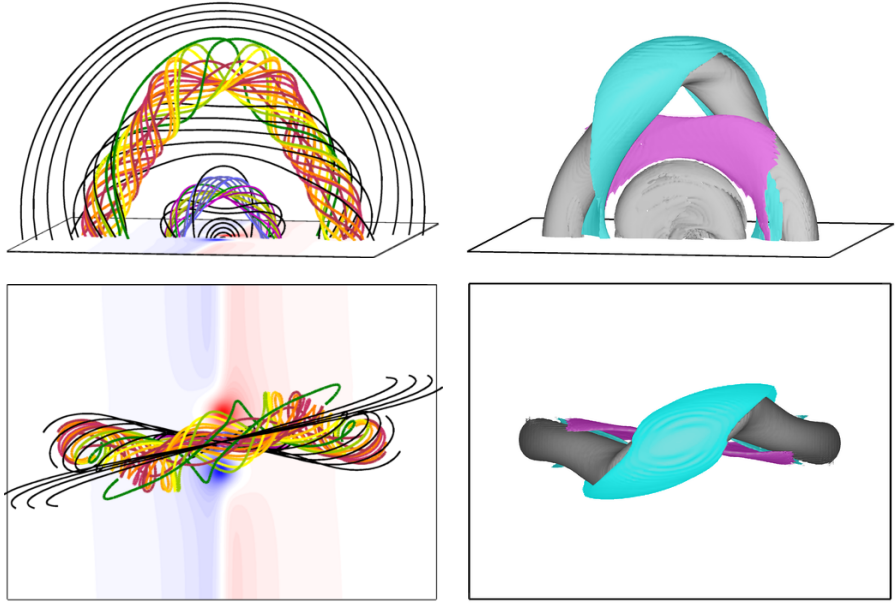}  
\caption{\label{f:flb_model}
 Partial eruption of the double flux rope equilibrium due to the
 instability of the upper rope after flux and current transfer from
 the lower rope (${\Delta}I_2/I_2=0.1$) near the marginal stability
 condition in the absence of flux transfer,
 $B_\mathrm{et}=1.7B_q\gtrsim B_\mathrm{et,\,cr}$. Field lines similar to
 Figure~\ref{f:relax} and current density isosurfaces at
 $|{\bf J}|=0.07J_\mathrm{max}$ show snapshots of the system at
 $t=645~\tau_\mathrm{A}$. The helical and vertical current sheets are
 colored in cyan and magneta, respectively. The ensuing reconnection
 of the unstable rope with the overlying field, which subsequently
 cuts the whole rope so that it remains confined in spite of the
 instability, is indicated by the dark green field lines.}
\end{figure*}

\begin{figure}
\centering \includegraphics[width=.85\linewidth]{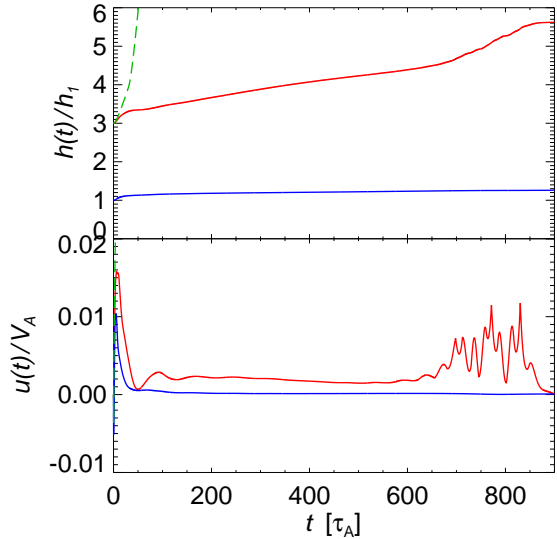}    
\caption{\label{f:rise_model}
 Height and velocity of the fluid elements at the apex points of the
 two ropes in the partial eruption shown in Figure~\ref{f:flb_model},
 which does not employ any velocity perturbation.
 The dashed green lines show the rise of the unstable upper flux rope
 when $B_\mathrm{et}$ is reduced by a factor 10.}
\end{figure}


The observations analyzed in Paper~I demonstrate that the most
significant change in the energy buildup phase prior to the eruption
consisted in the transfer of material and (necessarily
current-carrying) flux from the lower to the upper branch of the
filament (i.e., in the corona). To model this evolution, we next raise
the current $I_2$ through the upper flux rope and decrease $I_1$ in
the lower rope, keeping the total current $I_1+I_2$ at the equilibrium
value (and also $B_\mathrm{et}=1.7B_q$). As a consequence, the HFT
between the ropes moves a little bit downward, so that the cross
sections and the total flux in the ropes experience a change similar
to the currents. We have raised $I_2$ in small steps of $0.01I_2$
(with the corresponding changes of $I_1$ being close to
$-{\Delta}I_2/2$) and found indications for the onset of instability
of the upper rope at ${\Delta}I_2=0.04I_2$. For smaller changes of the
currents both flux ropes relax very near their initial positions. For
larger changes up to about ${\Delta}I_2=0.2I_2$, the lower flux rope
still relaxes near its initial position, while the upper rope erupts.

This is shown in Figures~\ref{f:flb_model} and \ref{f:rise_model} for
the case ${\Delta}I_2=0.1I_2$. The upper rope FR2 first moves
relatively quickly upward, to find the equilibrium position
corresponding to the increased current $I_2+{\Delta}I_2$ in about
$30~\tau_\mathrm{A}$. From this position, an approximately linear
ascent to $\approx\!1.5h_2$ commences. In the course of this rise, the
left-handed rope writhes into a projected forward S shape, piling up a
helical current sheet at its front side, analogous to the runs shown
in Figures~\ref{f:FR1up} and \ref{f:FR1down}. Simultaneously, the HFT
between the ropes collapses into a vertical current sheet. The upward
outflow resulting from the onset of reconnection in this current sheet
accelerates FR2 to a rise faster than linear
($t\gtrsim650~\tau_\mathrm{A}$). Since the overlying field resists the
rise, primarily due to the relatively strong $B_\mathrm{et}$, the top
part of the helical current sheet is quickly steepened and
reconnection between FR2 and the overlying field commences at the rope
apex, cutting the rope in two parts which remain confined. The lower
flux rope FR1 is stabilized primarily by the external toroidal field,
staying near the initial position for more than $10^3~\tau_\mathrm{A}$
with velocities remaining smaller than the rise velocity of FR2 by a
factor $\sim20$. The terminal height of FR2 in
Figure~\ref{f:rise_model} is the result of reconnection with the
overlying field. Overall the same behavior is found in the range
$0.04\le{\Delta}I_2/I_2\lesssim0.2$, with the velocities increasing
and reconnection commencing earlier for increasing ${\Delta}I_2$.



To demonstrate that the rise of FR2 is driven by an instability, the
run with ${\Delta}I_2/I_2=0.1$ is repeated with $B_\mathrm{et}$
reduced by a factor 10. Now the rise from the equilibrium position is
initially relatively close to an exponential function, as expected for
an instability developing in a weakly perturbed equilibrium (see
Figure~\ref{f:rise_model}). (As discussed above, for this low value of
$B_\mathrm{et}$ the lower flux rope FR1 is unstable as well.)

On the Sun, both the long-term stability of the double-decker filament
and the ejective eruption of the upper branch into a CME can be
allowed by an external toroidal field of the strength required for
stability at the position of the filament, but falling off with height
above the filament much faster than the model field. Our model field
is unrealistic in this regard but the coronal field is generally
expected to satisfy this condition (see, e.g., the modeling of an
active region containing a filament in \citealt{Su&al2011} and
\citealt{Kliem&al2013}).

The transfer of flux from the lower to the upper flux rope is
different from flux transfer by reconnection at the HFT between the
ropes, e.g., tether-cutting reconnection. Such reconnection exchanges
flux in both ropes with ambient flux. It is conceivable that the flux
transfer is enforced by configuration changes of the current in the
lower flux rope, which, in turn, can be enforced by changes in the
photospheric boundary. It is well known that current-carrying flux
rises if it is stressed by photospheric motions
\cite[e.g.,][]{Mikic&Linker1994,tk03}. In the event considered in
Paper~I, the dominant photospheric change prior to the eruption
consisted in the ongoing dispersion of the diffuse positive flux in
the northern hook of the filament, which may have enforced an exchange
of flux between the two filament branches. At the southern end of the
filament, the moat flow of the negative sunspot moved flux in the
direction from the end point of the lower branch to the end point of
the upper branch (see Figure~2 in Paper~I).

We have attempted to also model such driving by prescribing changes of
the configuration only in the photospheric boundary. From each
footpoint region of the lower rope, a patch of current-carrying flux
was moved to the neighboring footpoint region of the upper rope by
prescribing appropriate horizontal motions in the photosphere
(satisfying symmetry with respect to the point under the flux rope
apex). Although a number of different geometries for the path of the
flux patch and different amounts of transferred flux were considered,
none of the experiments succeeded in yielding a partial eruption as
observed in the events analyzed in Paper~I and
\citet{Zhu&Alexander2014}. The perturbation always made both ropes
unstable, with various combinations of upward and downward
displacements of the ropes. FR1 was destroyed when moving downward and
merged with FR2 when moving upward; the remaining/merged upper rope
erupted in some runs and relaxed to an equilibrium at the new height
otherwise. It appears that the generalized double Titov-D\'emoulin
equilibrium is not appropriate for the modeling of such directly
driven flux transfer between the ropes in their slow-rise phase
because the ropes are too strongly twisted in their outer part. (The
twist profile in the TD99 model increases toward the surface of the
current channel in the rope; see Figure~2 in \citealt{tkt04}.) The
flux bundle rooted in the moving photospheric flux patches, which
originate from the surface of the lower current channel, winds
considerably about the axis of the lower rope, so that the rope is
always substantially perturbed. A less twisted, numerically
constructed double flux rope equilibrium, like the one described in
the following section or one obtained through the flux rope insertion
method \cite[e.g.,][]{Su&al2011}, may allow the modeling of flux
transfer driven at the photosphere with the lower rope remaining
stable.

\section{Splitting Flux Bundle in the Slow Rise Phase of a CME}
\label{s:splitting_FR}

The formation and partial eruption of a double flux rope configuration
was also found in an MHD simulation that was designed to model the
well-known filament eruption and CME on 1997 May 12
\citep[e.g.,][]{thompson98,webb00}. The details of this simulation
will be described elsewhere (Linker et al., in preparation); here we
merely summarize its main features and focus on the evolution shortly
before the eruption.

\begin{figure*}[t]
\centering \includegraphics[width=\textwidth]{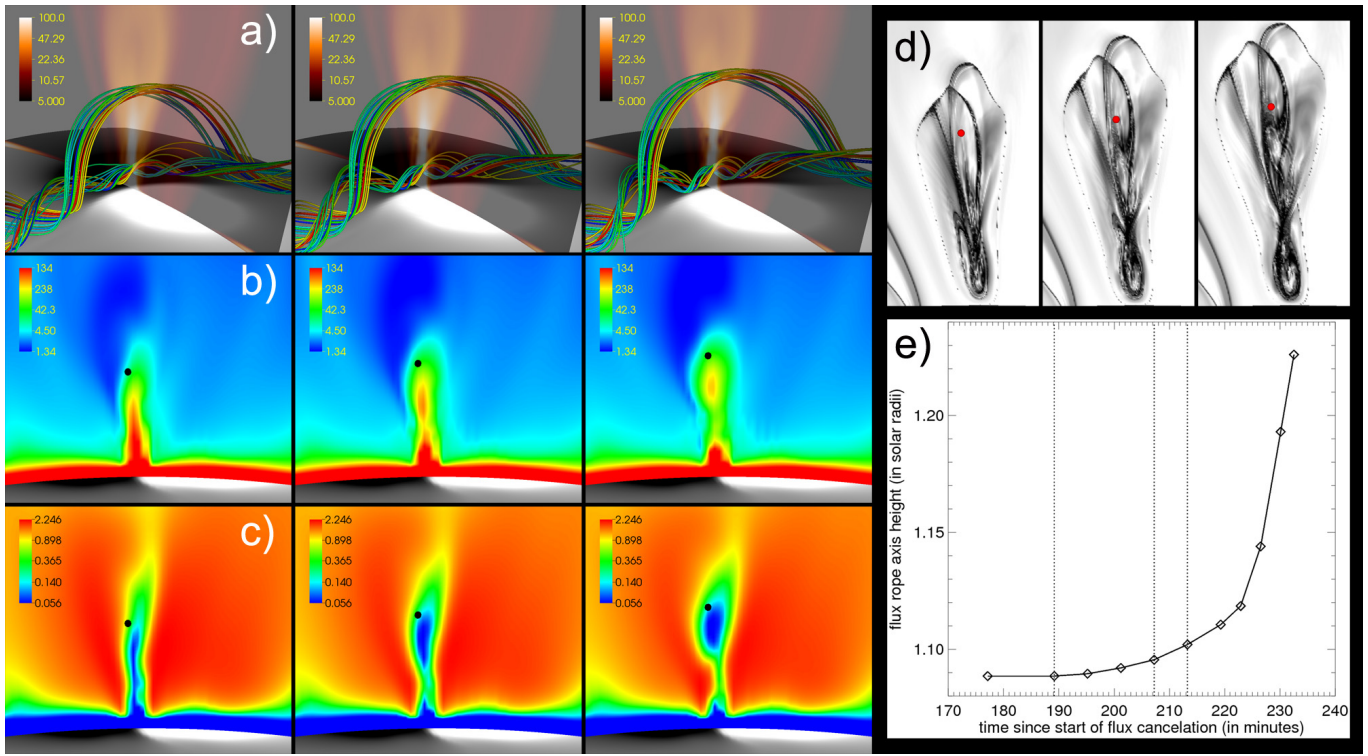} 
   \caption {Snapshots of the 1997 May 12 simulation during the pre-eruptive phase. Panels
(a--d): Vertical cuts perpendicular to the flux rope axis, taken at the apex of the upper rope;
the specific times of the sub-panels from left to right are marked by dotted lines in the
time-height profile of the upper flux rope axis apex shown in Panel (e). Panel (a) shows the
(transparent) logarithmic distribution of the ratio of electric current density and magnetic
field strength, $j/B$ (in normalized units), and magnetic field lines that outline the cores of
the upper and lower flux rope in an oblique view. Panels (b--d) are views along the upper flux
rope axis, showing mass density $\rho$ (in $10^{-16}$ g\,cm$^{-3}$), temperature $T$ (in MK), and
the logarithmic distribution of the squashing factor $Q$ (shown in a zoomed view),
respectively. The dots in panels (b--d) mark the approximative position of the upper flux rope
axis apex. The radial magnetic field is shown in panels (a--c) at the photospheric plane, where
white (black) colors outline strong positive (negative) flux.
%
    \label{fig:may97} }
\end{figure*}

The simulation code solves the standard resistive and viscous MHD
equations on a spherical grid and incorporates radiative losses,
thermal conduction parallel to the magnetic field, and an empirical
coronal heating function \citep[see, e.g.,][]{mikic99,lionello09}. The
initial magnetic field in the simulation is obtained from a potential
field extrapolation, using a combination of synoptic and real-time
line-of-sight magnetograms \citep[for a simplified model of the
large-scale magnetic configuration around the time of the eruption,
see][]{titov08}. After a solar wind MHD solution is obtained by
relaxing the system to a steady state, the CME-producing active region
is energized by a combination of photospheric shear flows and
transverse field ``emergence'', which both keep the radial flux
distribution unchanged \citep[e.g.,][]{linker01,bisi10}. This produces
a highly sheared core field with little indications of twist, i.e., of
flux rope geometry. Finally, the system is further energized, and an
eruption is triggered, via flux cancellation, driven by localized
flows converging toward the photospheric polarity inversion line (PIL)
\citep[e.g.,][]{linker03}.

During the energization of the system, a sheet-like, coherent
structure of cold and dense plasma forms within the core field
(Figure~\ref{fig:may97}(b),(c)). Interestingly, a significant fraction
of this plasma is not located in concave-up field line segments, so
the common picture of prominence material held against gravity in
dipped fields does not seem to fully apply here. The detailed
mechanisms by which this ``prominence'' is formed and maintained in
the simulation require further study; both levitation of chromospheric
plasma and condensation of coronal material may be involved. For our
present purpose, it is sufficient to consider its evolution in the
phase leading up to the eruption.

It has been shown that magnetic reconnection associated with flux
cancellation successively transforms a sheared arcade into a flux
rope, which slowly detaches from the photosphere, leaving behind short
arched field lines \citep[e.g.,][]{vbm89,amari03,aulanier10}. A
similar process occurs in our simulation during the cancellation
phase, although the evolution is more complex (Titov et al., in
preparation). As can be inferred from vertical cuts of the so-called
squashing factor $Q$ \citep{titov02}, the core field consists of
several flux bundles before it erupts (Figure~\ref{fig:may97}(d)). The
central flux bundle contains two adjacent flux ropes: a highly twisted
low-lying one and an arched, less twisted upper one
(Figure~\ref{fig:may97}(a)). These two structures start to form
relatively early in the cancellation phase, and gradually develop a
flux rope geometry as they accumulate twist about their respective
axes. However, in contrast to the distinct flux ropes considered in
the previous subsection, a clear boundary between them does not
develop; rather they remain merged to some extent for most of the
cancellation phase. (Interestingly, although the upper flux rope
possesses twist, it does not contain dipped field lines below its
axis, which is due to its strong curvature.)

In the course of the flux cancellation and associated reconnection,
the core field first expands quasi-statically until, after about 3.5
hours, its slow evolution transitions into a fast rise phase, marking
the onset of the eruption (Figure~\ref{fig:may97}(e)). In the
pre-eruptive phase, the central flux bundle becomes increasingly
stretched in the vertical direction. Shortly before the eruption, the
splitting into two parts becomes more pronounced
(Figure~\ref{fig:may97}(d)), which is associated with tether-cutting
reconnection of ambient flux into the core flux at the HFT between the
two main flux bundles. The added flux runs under the apex of the upper
flux rope and contributes to its twist, thus, it potentially acts
destabilizing. For the lower rope, the added flux acts like
strengthened overlying flux, thus stabilizing. The increasing
separation of the flux ropes is accompanied by a splitting of the
plasma sheet (Figure~\ref{fig:may97}(b),(c)). Subsequently, the
eruption carries away the upper flux rope and the top part of the
plasma sheet, while the lower rope and the bottom part of the sheet
remain at low heights.

The origin of the splitting of the central flux bundle must be
different from the partial eruption mechanism modeled by \cite{gf06}.
There, the splitting of one coherent flux rope into two parts was
associated with reconnection occurring in a current layer that formed
in the course of the rope's eruption, due to a deformation (writhing)
of the rope by the helical kink instability. In our case, the
splitting occurs already in the pre-eruptive stage, during which no
noticeable helical deformation takes place
(Figure~\ref{fig:may97}(a)). Moreover, two flux ropes form within the
central flux bundle even earlier in the evolution. In a later
simulation described in \cite{fan10}, where the eruption was driven by
the torus instability, a current layer formed within the flux rope
already in the pre-eruptive phase, and a similar breaking of the
magnetic structure as in our simulation was observed (Y. Fan, private
communication). Further study is required to fully understand how the
splitting of the magnetic flux occurs in our simulation. A number of
aspects may play a  role, as for example the complexity of the initial
potential field, the specific techniques by which the system is
driven, gravity in regions of strong density, and reconnection at bald
patches in the outer parts of the PIL.

By producing two stacked flux ropes early on in the evolution leading
up to an eruption, our simulation supports the conjecture that such a
configuration can exist in stable state for long periods in the
corona, and it provides indications for their possible formation.
Also, the splitting of the plasma sheet corresponds nicely to the
observed separation of the two filament branches in the event analyzed
in Paper~I. However, the plasma splitting occurs only rather shortly
before the eruption, different from the 2010 August~7 event (although
the separation of the branches was most pronounced shortly before the
eruption of the upper branch; see Figure~4 in Paper~I). We did not
succeed in producing a stable or longer-lasting configuration with a
split plasma sheet by, for instance, switching off the flux
cancellation at earlier times in the simulation. It seems that, at
least for the specific parameters that were used to control the
formation and driving of the core field in our simulation, such a
configuration is difficult to obtain. The absence of plasma splitting
at an earlier stage of the simulation might be due to the absence of
field line dips in the upper flux rope. It remains to be seen whether
the formation of (meta-)stable double-decker filament configurations
can be modeled by simulations similar to the one presented here, or
whether different physical mechanisms (as for example flux emergence)
are required.

\section{Conclusions} \label{s:concl}

Stimulated by indications that vertically split (``double-decker'')
filaments and filament-sigmoid systems may form in double flux rope
equilibria \cite[Paper~I;][]{Zhu&Alexander2014, XCheng&al2014}, the
present investigation shows that an approximate analytical equilibrium
of two concentric, toroidal, force-free flux ropes can be constructed
by a generalization of the methods developed in \citet{td99}. The
technique can be used for ratios of the major torus radii
$R_2/R_1\gtrsim2.5$ and should be supplemented by numerical MHD
relaxation to a nearby numerical equilibrium in the range
$R_2/R_1\lesssim4$.

The equilibrium is stabilized by an external toroidal field
$B_\mathrm{et}$ of sufficient strength, which can be considerable if
the flux ropes are located relatively close to each other. For the
geometry studied here, $R_2/R_1=2.5$ and the ratio of minor torus
radii $a_2/a_1=1.5$, we find that
$B_\mathrm{et}>B_\mathrm{et,\,cr}\approx1.7B_q$ is required, where
$B_q$ is the external poloidal field strength at the position of the
inner (lower) flux rope.

The analytical construction of a double flux rope equilibrium with
$B_\mathrm{et}\ne0$ in this paper is restricted to the case that a
line current at the symmetry axis of the tori is the source of
$B_\mathrm{et}$. Therefore, $B_\mathrm{et}$ decreases only weakly with
increasing distance $R$ from the symmetry axis,
$B_\mathrm{et}\propto{R}^{-1}$. This allows only confined eruptions to
be modeled (except in the case that $B_\mathrm{et}$ is set
considerably below $B_\mathrm{et,\,cr}$).

If the external toroidal field strength is reduced, then both flux
ropes tend to become torus unstable. Typically the lower (inner) flux
rope exhibits the stronger instability. This is due to the facts
(\textit{i}) that the poloidal field due to the upper (outer) flux
rope is oppositely directed to the external poloidal field at the
position of the lower flux rope, giving the total poloidal field at
this position a steep decrease with height, and (\textit{ii}) that
field and current are generally stronger in the lower flux rope. If
the lower flux rope erupts upward, it pushes the upper rope upward as
well and merges with it. A full eruption of the configuration results
which is quite similar to the eruption of a single flux rope. If the
lower rope moves downward, it reconnects with the ambient field,
splitting and coming to rest low in the box. This is accompanied by
the upward eruption of the upper rope. However, it is also possible
that only the upper flux rope turns unstable, with the lower rope
staying in place without experiencing any significant change. This
occurs in a scenario suggested by the observations in Paper~I and
\citet{Zhu&Alexander2014} and demonstrated here: transfer of flux and
current from the lower to the upper flux rope.

An equilibrium consisting of two force-free flux ropes, arranged
vertically above a photospheric boundary, is also numerically obtained
through an evolutionary sequence of shear flows, flux emergence, and
flux cancellation in the photosphere. The stable double flux rope
structure shows a slow rise in the cancellation phase. This evolution
gradually transitions into a faster rise involving tether-cutting
reconnection with the ambient field at the HFT between the ropes. Such
reconnection lowers the stability of the upper flux rope and improves
the lower rope's stability. Thus, it is a second potential driver for
a partial eruption. The simulation indeed yields an eruption of the
upper flux rope while the lower rope remains at low heights.

A third possibility, although less likely, consists in building up
supercritical twist for onset of the helical kink instability only in
the upper flux rope.

\section{Discussion} \label{s:discussion}

Both the existence of stable double flux rope equilibria and the
possibility that only the upper flux rope loses stability support the
suggestion that the split, partially erupting filaments investigated
in Paper~I and \citet{Zhu&Alexander2014}, and the filament-sigmoid
systems investigated in \citet{liu10} and \citet{XCheng&al2014}, may
have formed in such a configuration.

This topology may not be a rare occurrence, since split filaments and
prominences are seen quite frequently and since partial eruptions are
not uncommon \cite[e.g.,][]{pevtsov02}. In hindsight rather many
filament-sigmoid systems may be compatible with a double flux rope
equilibrium, for example, the one in AR~10944, which partially erupted
on 2007 March~2 \citep{Sterling&al2007, liu08}. The change of the
originally discontinuous EUV sigmoid into a single continuous
structure lying above the filament at the onset of this eruption is
consistent with the tether-cutting reconnection between the flux ropes
and the ambient field observed in the simulation in
Section~\ref{s:splitting_FR}. One can expect a more general relevance
also from the fact that the simulation in Section~\ref{s:splitting_FR}
was designed to model another event and yet shows a double flux rope.
The existence of two flux ropes is quite likely in an event on 2010
March~30 which exhibits both confined and ejective components in a
common eruption \citep{Koleva&al2012}. Both main branches of a split
prominence were seen to erupt successively in an event on 2010 April~8
\citep{Su&al2011}, possibly providing an example for the case that the
lower rope in a double flux rope equilibrium loses stability first.
Moreover, the double flux rope topology may not only be relevant for
split filaments and prominences, since one of the flux ropes,
especially the upper one, may be void of absorbing material.

The episodic transfer of mass and magnetic flux to the upper branch of
a split filament was found to be a likely destabilization mechanism
for the upper branch in two events \cite[Paper~I
and][]{Zhu&Alexander2014}. Recently, such a ``flux feeding'' was
observed to occur between chromospheric fibrils and an overlying
filament, destabilizing the filament in an eruption on 2012 October~22
\citep{QZhang&al2014}. Thus, the transfer of mass and magnetic flux to
a filament may be of more general relevance for the triggering of
eruptions.

The support for the double flux rope configuration given here does not
imply that the alternative configuration of a single flux rope
situated above a magnetic arcade is less feasible. That configuration
actually has advantages in terms of stability (it does not necessarily
require a strong external shear field component to suppress
instabilities), and it naturally allows partial eruptions that do not
strongly perturb the lower part of the flux. However, it has the
disadvantage that a basically different magnetic structure for the two
filament branches is implied.

A formation mechanism for the double flux rope configuration in
Figure~\ref{f:cartoon_w_HFT}(a) by the splitting of a coronal flux
bundle driven by photospheric flows and flux cancellation has been
demonstrated here (Section~\ref{s:splitting_FR}). A further mechanism
is given by the emergence of a flux rope under an existing flux rope,
as briefly discussed in Paper~I. Additionally, the configuration may
form in a generic manner in an extended period of flux cancellation,
which first forms the configuration of Figure~\ref{f:cartoon_w_HFT}(b)
and then produces the lower flux rope by acting on the arcade under
the HFT.

Several new scenarios for partial eruptions are suggested by the
investigations presented in Paper~I and here. The upper part of a
split flux system can erupt without strongly perturbing the lower part
in the double flux rope configuration when only the upper flux rope is
unstable (Figures~\ref{f:flb_model}--\ref{fig:may97}), which can be
achieved by flux transfer from the lower to the upper rope and by
tether-cutting reconnection with the ambient field in the space
between the ropes. Similar evolutions are possible in the
configuration with a single flux rope separated by an HFT from an
underlying arcade (Figure~\ref{f:cartoon_w_HFT}(b)). The double flux
rope configuration also allows the eruption of the upper flux rope
accompanied by downward motion and eventual destruction of the lower
flux rope if the lower flux rope is unstable. In each of these cases,
the flux can be split already during long periods before the eruption
starts, in line with the properties of the filaments studied in
Paper~I and \citet{Zhu&Alexander2014}. This distinguishes them from
the well known mechanism of a splitting unstable and line-tied flux
rope in \citet{gf06} \cite[see also][]{gilbert01}, which involves a
splitting of the flux only in the course of the eruption.

\acknowledgments B.K.\ acknowledges support by the DFG, the STFC,
the NSF (grant AGS-1249270) and by the Chinese Academy of Sciences
under grant 2012T1J0017.
The contributions of T.T., V.S.T., R.~Lionello, and J.A.L. were
supported by NASA's HTP, LWS, and SR\&T programs, and by CISM (an NSF
Science and Technology Center). Computational resources were provided
by NSF TACC in Austin and by NASA NAS at Ames Research Center.
R.~Liu, C.L., and H.W. were supported by NSF grants AGS-0849453,
AGS-0819662, and AGS-1153226.

\bibliographystyle{apj}
\bibliography{ribbon}

\end{document}